\documentclass[12pt,preprint]{aastex}

\shorttitle{Yellow and Red Supergiants in the LMC}

\shortauthors{Neugent et al.}

\begin{document}

\title{Yellow and Red Supergiants\\ in the Large Magellanic Cloud}

\author{Kathryn F. Neugent\altaffilmark{1}, Philip Massey\altaffilmark{1} and Brian Skiff}
\affil{Lowell Observatory, 1400 W Mars Hill Road, Flagstaff, AZ 86001;\\ kneugent@lowell.edu; phil.massey@lowell.edu; bas@lowell.edu}

\author{Georges Meynet}
\affil{Geneva University, Geneva Observatory, CH-1290 Versoix, Switzerland; georges.meynet@unige.ch}

\altaffiltext{1}{Visiting astronomer, Cerro Tololo Inter-American Observatory (CTIO), a division of the National Optical Astronomy Observatory, which is operated by the Association of Universities for Research in Astronomy, Inc., under cooperative agreement with the National Science Foundation.}

\begin{abstract}
Due to their transitionary nature, yellow supergiants provide a critical challenge for evolutionary modeling. Previous studies within M31 and the SMC show that the Geneva evolutionary models do a poor job at predicting the lifetimes of these short-lived stars. Here we extend this study to the LMC while also investigating the galaxy's red supergiant content. This task is complicated by contamination by Galactic foreground stars that color and magnitude criteria alone cannot weed out. Therefore, we use proper motions and the LMC's large systemic radial velocity ($\sim$278 km s$^{-1}$) to separate out these foreground dwarfs. After observing nearly 2,000 stars, we identified 317 probable yellow supergiants, 6 possible yellow supergiants and 505 probable red supergiants. Foreground contamination of our yellow supergiant sample was $\sim$80\%, while that of the the red supergiant sample was only 3\%. By placing the yellow supergiants on the H-R diagram and comparing them against the evolutionary tracks, we find that new Geneva evolutionary models do an exemplary job at predicting both the locations and the lifetimes of these transitory objects. 
\end{abstract}


\section{Introduction}
\label{intro}
As a massive star reaches the end of its life, it eventually exhausts its hydrogen supply and begins burning helium, a phase that lasts for only $\sim$10\% of the massive star's lifetime. In many cases, it has now become a mid- to late-type supergiant. In this paper we focus on the cooler of these supergiants within the Large Magellanic Cloud (LMC): the yellow supergiants (YSGs) and the red supergiants (RSGs). 

In comparison to the $\sim$3,000 un-evolved OB stars more massive than 20$M_\odot$ in the Small Magellanic Cloud (SMC), Neugent et al.\ (2010) found only 176 YSGs, a number complete to a few percent. This rarity is due to the transitionary nature of YSGs: they represent a short phase within a massive star's life as it passes from the blue side of the Hertzsprung-Russell diagram (HRD) to the red supergiant stage or from the red back to the blue. However, their rarity only boosts their importance when testing current stellar evolutionary theory since the evolutionary model's ability to predict the locations and numbers of these short-lived stellar objects on the HRD provides a crucial test. Previous studies of YSGs in the SMC (Neugent et al.\ 2010) and in M31 (Drout et al.\ 2009) show that the Meynet \& Maeder (2005) Geneva evolutionary models overestimate the lifetimes of YSGs by large factors. Reliable evolutionary tracks affect not only the studies of massive stars, but the usefulness of population synthesis codes such as STARBURST99 (Leitherer et al.\ 1999, Vazquez \& Leitherer 2005), used to interpret the spectra of distant galaxies.

\begin{figure}
\plotone{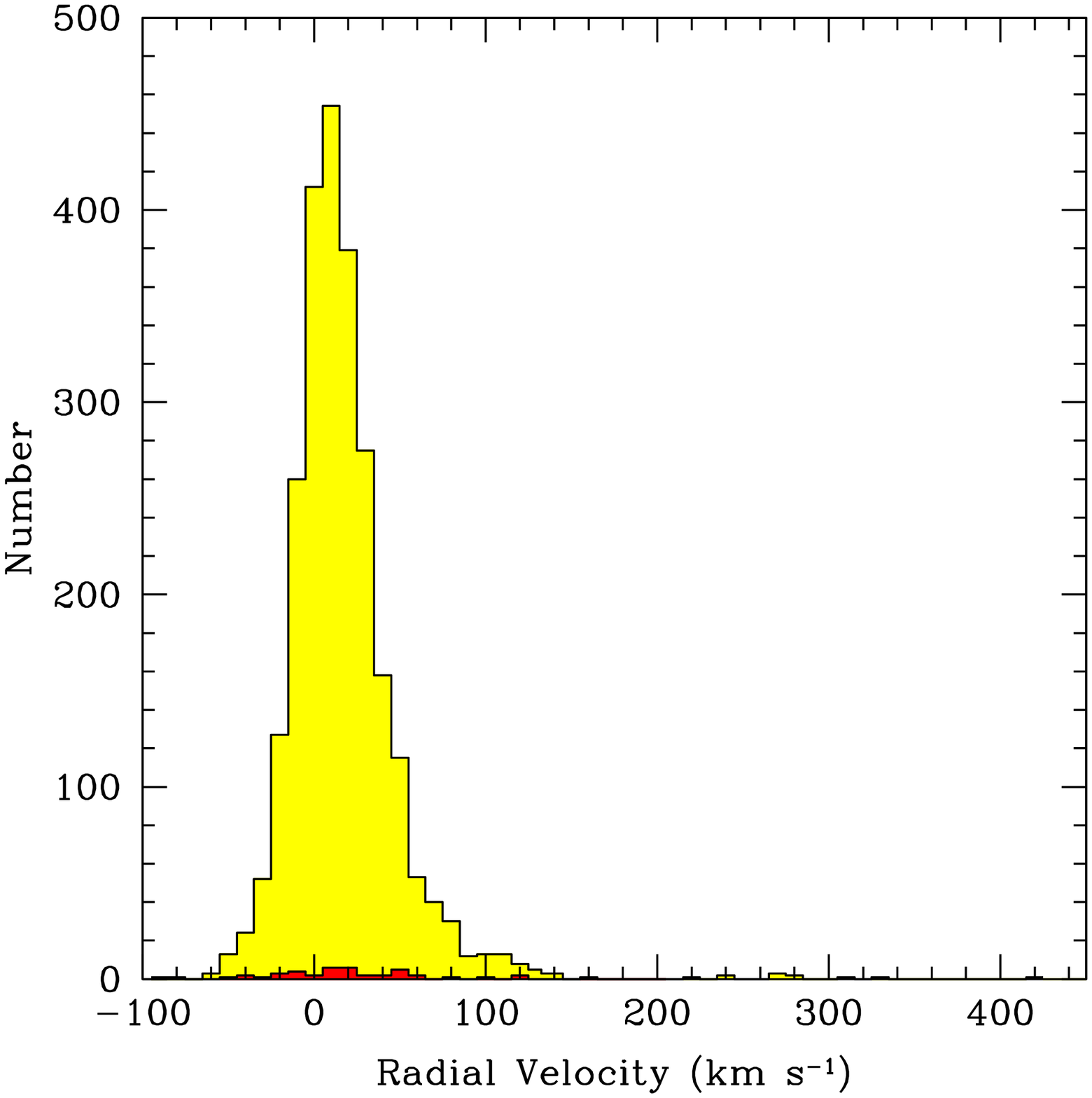}
\caption{\label{fig:besAll} Predicted LMC foreground contamination. Contamination for both YSGs (yellow) and RSGs (red) was estimated using the Besan\c{c}on models. The models were run using the same color ranges and proper
motion selection as our candidates and covering the same area of sky.}
\end{figure}

Testing the evolutionary models by identifying a complete sample of supergiants within the LMC is complicated by foreground dwarf contamination. While this issue is pertinent when identifying both yellow and red LMC supergiants, the number of foreground stars in the appropriate color and magnitude range is much greater for LMC YSGs than for RSGs. This is because a red dwarf would have to be very close by to be in the same magnitude range, and the majority of these near-by dwarfs can be eliminated using proper motion cut-offs. Figure~\ref{fig:besAll} shows the expected LMC foreground contamination primarily by yellow and red disk dwarfs as predicted by the Besan\c{c}on Milky Way models (Robin et al.\ 2003) using the same area, proper motion cutoff and magnitude and color criteria as we use when defining our sample (see \S2). Since evolutionary considerations suggest we will find many more RSGs than YSGs within the LMC, the contamination of YSGs will be much greater. In fact, Massey \& Olsen (2003) found their RSG foreground contamination (without a proper motion cut-off) to be 11\% in the SMC, while Neugent et al.\ (2010) found a contamination closer to 65\% when looking at SMC YSGs. We plan to circumvent this issue by determining the radial velocities of all of our observed supergiant candidates since there will be minimal overlap between the radial velocities of LMC and Milky Way members. Further details will be discussed in \S3.

After identifying a complete sample of LMC YSGs, we will be able to test the Geneva evolutionary models observationally by comparing the relative numbers of supergiants as a function of luminosity. We've done this for the low metallicity SMC (log$\frac{O}{H} + 12 = 8.1$; Russell \& Dopita 1990) and for the high metallicity M31 (log$\frac{O}{H} + 12 = 9.1$; Zaritsky et al.\ 1994) and here we extend this study to RSGs while investigating the LMC's intermediate metallicity of log$\frac{O}{H} + 12 = 8.4$ (Russell \& Dopita 1990). In addition, we have at our disposal a new set of Geneva evolutionary models to test. 

We begin this paper in \S2 by describing our observation and reduction procedures. In \S3 we discuss how we separated foreground stars from our LMC yellow and red supergiants and determined LMC membership. In \S4 we examine both our contamination and completeness before, in \S5, putting the stars on the HRD and testing the current Geneva evolutionary models. Finally, in \S6 we summarize our findings and describe future goals.

\section{Selections, Observations and Reductions}
\subsection{RSG and YSG Candidate Selections}
\label{sel}
The YSG and RSG LMC candidates were initially selected using the USNO CCD Astrograph Cagalogue Part 3 (UCAC3). We first chose stars within the LMC's visible disk by including those within a $3\fdg5$ radius centered at $5^{h}15^{m}20^{s}$ $-69^\circ20\arcmin10\arcsec$ (J2000). We then attempted to weed out foreground stars by excluding those with absolute proper motion values greater than 15 mas year$^{-1}$ in $\alpha$ or $\delta$. Finally, we used the UCAC3 quality codes to remove possible galaxies, clusters and double stars.

\begin{figure}
\plotone{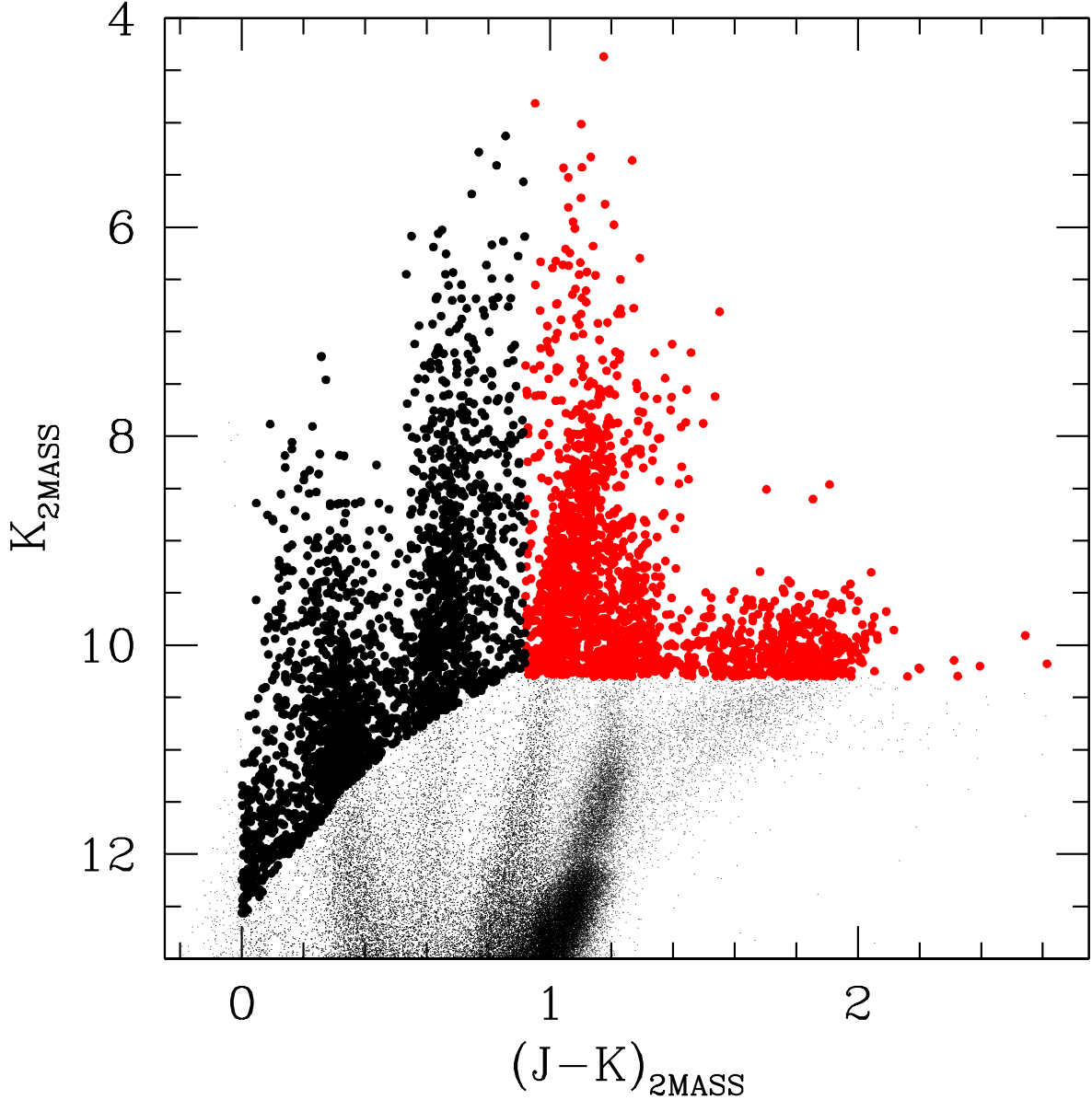}
\caption{\label{fig:JMKvsK} Color and magnitude selection criteria for the YSG and RSG candidates. The tiny black dots represent all possible targets before any selection based on color and magnitude. The YSG candidates (black filled circles) extend between $(J-K)_{\rm 2MASS} > 0$ when $K_{\rm 2MASS} = 12.6$ and $(J-K)_{\rm 2MASS} < 0.9$ when $K_{\rm 2MASS} = 10.2$ and extend redwards. The RSG candidates (red filled circles) pick up at $(J-K)_{\rm 2MASS} > 0.9$ when $K_{\rm 2MASS} = 10.2$. We believe the reddest colors to be caused by high reddening and/or confusion and expect to find the majority of the RSGs within $0.9 < (J-K)_{\rm 2MASS} < 1.2$, or a $T_{\rm eff}$ range of $3500 - 4500$ K.}
\end{figure}

To select a sample with appropriate magnitude and color ranges for both YSGs and RSGs, we relied on the stars' 2MASS photometry (Skrutskie et al.\ 2006). To select YSGs, we followed the procedure described in Neugent et al.\ (2010) for the SMC. Our goal was to be complete down to $12M_\odot$ after allowing for the different distance to the LMC. Using a $T_{\rm eff}$ range of $4800 - 7500$ K, we used the (older) Geneva evolutionary tracks (Maeder \& Meynet 2001) and the $J-K$ colors of Kurucz's (1992) ATLAS9 atmosphere models to define $K$ magnitude limits as a function of $J-K$ for a $12M_\odot$ YSG. When selecting RSGs, we were concerned about potential contamination by intermediate-mass asymptotic giant branch stars since they overlap in luminosity with RSGs (see Brunish et al.\ 1986). Therefore, we weren't complete down to as low a mass for the RSGs as we were for the YSGs and simply defined a flat $K$ magnitude cut-off. The color selection criteria of both the YSGs and RSGs candidates is shown in the color-magnitude (CMD) Figure~\ref{fig:JMKvsK}. After this, we ended up with 2187 YSG candidates and 1949 RSG candidates.

\subsection{Observations}
All data were collected using the Cerro Tololo 4-meter telescope and Hydra, a 138 fiber multi-object spectrometer with a 2/3$^\circ$ field of view. Before observing, we matched the 2$\arcsec$ diameter fibers to specific YSG and RSG candidates. Higher priority was given to the YSG candidates since foreground contamination is more likely within their magnitude and color ranges. Additionally, fields with the most new targets were assigned higher priorities so if (when) bad weather struck or a mechanical glitch occurred, it was clear which fields should be observed first.

Observations spanned over a clear eight night run covering (UT) 2011 January 18-25. (In addition, we used data collected in 2009 October described by Neugent et al.\ 2010). We used the KPGL-D grating and OG 515 blocking filter in order to observe the 7700-9200~\AA\ region, centered on the Ca II triplet ($\lambda\lambda 8498, 8542, 8662$), that we intended to use for radial velocities. The set up was the same as that used for the SMC YSG project (Neugent et al.\ 2010). This achieved a spectral resolution of 2.6 \AA\ (3 binned pixels). Each field was observed for three consecutive exposures of five minutes each, followed by a short exposure of a HeNeAr lamp for wavelength calibration, and a projector lamp exposure obtained for flat fielding. We also obtained dome flat exposures during the late afternoons on some days with the fibers configured into ``a great circle". Additionally, a series of bias exposures were obtained each night. We observed five Geneva radial-velocity standard stars (HD115521 - M2III, HD42807 - G2V, HD6655 - F8V, HD83516 - K0III, HD84441 - G0III) for use as cross-correlation templates. These stars were observed on multiple occasions during the run, with at least several standards observed each night for a total of 24 exposures. 

\begin{figure}
\plotone{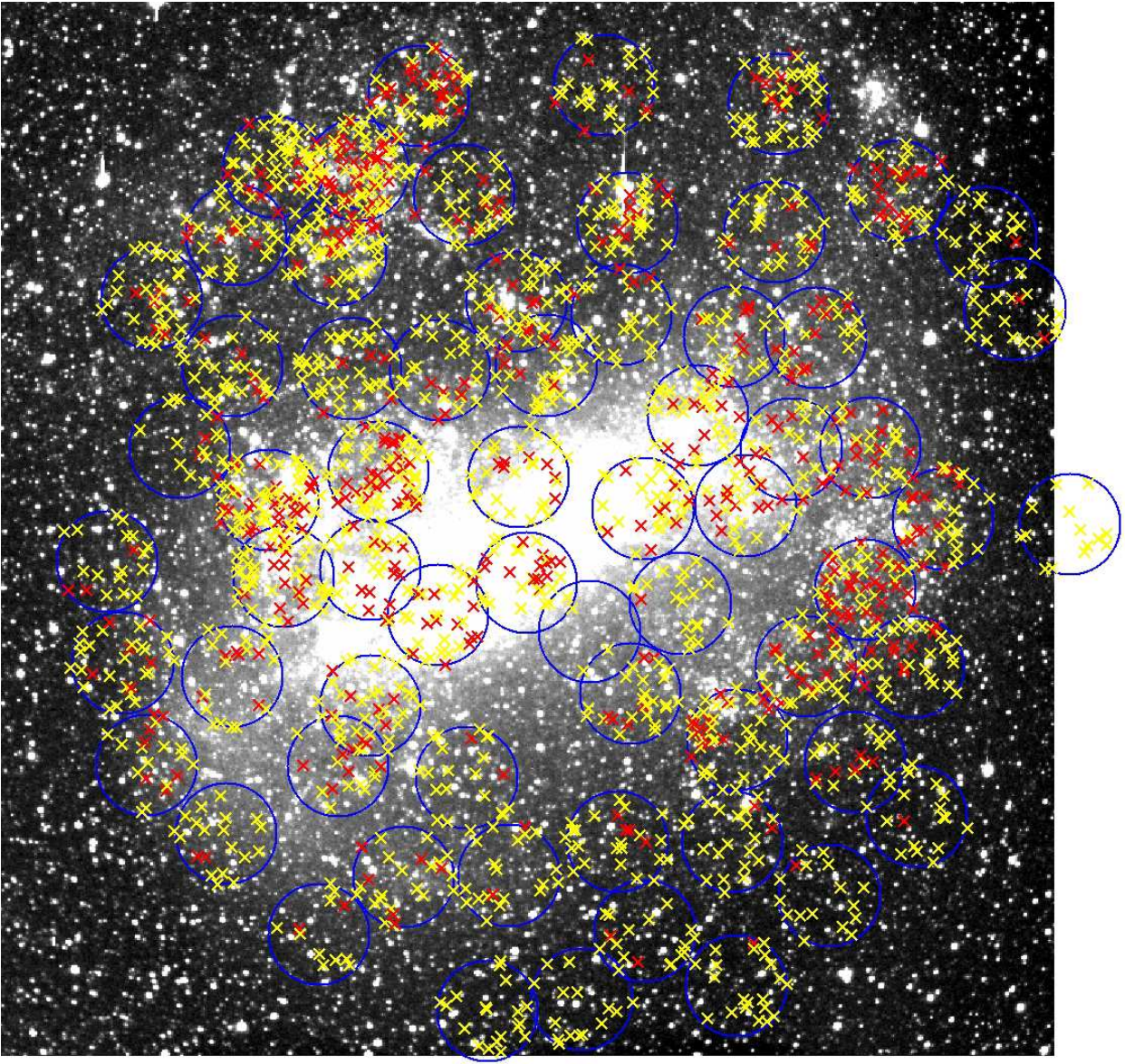}
\caption{\label{fig:config} Locations of observed fields and candidates. The blue 2/3$^\circ$ circles represent the 64 fields observed while the yellow $\times$s represent the YSG candidates and the red $\times$s represent the RSG candidates observed. The background image was obtained using the ``parking lot" camera (Bothun \& Thompson 1988).}
\end{figure}

Overall, we observed 64 fields for a total of 1528 (70\%) unique YSG candidates and 865 (44\%) unique RSG candidates. Additionally, four fields were observed twice and a few stars were assigned to multiple fields providing us with 156 YSG candidates and 23 RSG candidates observed more than once. Figure~\ref{fig:config} shows the locations of the observed fields and candidates.

\subsection{Reductions}
We overscan-subtracted and trimmed our data by first removing a scalar value determined from the overscan columns, and then trimming off the portion of the image containing the overscan columns. We additionally removed any left-over bias structure by subtracting an averaged bias. The IRAF\footnote{IRAF is distributed by the National Optical Astronomy Observatory, which is operated by the Association of Universities for Research in Astronomy (AURA) under cooperative agreement with the National Science Foundation.} task ``dohydra" was then used to extract, flat-field, and wavelength calibrate the spectra. After some investigation we decided to use the dome flats as our flat-fields rather than the projector flats. Although either would remove the pixel-to-pixel variations, the dome flats did a better job of removing the fiber-to-fiber variations, as the illumination of the projector flats was much less uniform. Sky subtraction was provided by preselected ``sky" fibers which were then averaged for each field after removing highly deviant results. Finally, the three spectra for each star were combined after rejecting bad pixels using IRAF's ``avsigclip" algorithm.

After examining the spectra, we found that 69 (4\%) of the candidate YSGs and 343 (59\%) of candidate RSGs we assigned to fibers yielded nothing but sky. While 4\% is understandable (a few fibers had low transmission), 59\% is much larger than expected. However, plotting the $J-K$ values of these 343 ``invisible" candidate RSGs against $K$ immediately yielded the explanation. RSGs very rarely have a $T_{\rm eff}$ lower than 3500 K, which corresponds to $J-K = 1.2$. But a glance at Figure~\ref{fig:JMKvsK} shows that many of our candidate RSGs had $J-K$ values much larger than 1.2. We were gratified to discover that most of our ``invisible" RSGs (93\%) fell in this region of large $J-K$ values. These cannot simply be spurious sources, as they have both measured proper motions from the UCAC3 {\it and} 2MASS photometry.  Indeed, as we discuss in \S5.2, there are many stars in this region of the CMD for which we did obtain spectroscopy and confirm membership; these show strong evidence of being highly dusty objects, deserving future followup\footnote{Note that the fainter and more heavily reddened of these will indeed be undetectable in our exposures, which are basically in the $I$ band. Consider a RSG with $T_{\rm eff}=3800$ K (typical of M0 I's) and hence intrinsic colors of $(J-K)_0 = 1.0$, $(V-K)_0=4.0$, $(V-I)_0=1.8$ (Levesque et al.\  2006), and $V_0=13.5$.  With the normal amount of LMC reddening, E(B-V)=0.13 (Massey et al.\ 1995), we expect the star to have $V\sim 14.0, K\sim 9.6$, and an $I$ magnitude of about $12.0$.  However, if the star has a circumstellar dust shell resulting in $J-K=1.8$, then $E(J-K)=0.8$ and hence $A_V=4.9$~mag, assuming the dust follows a normal reddening law (i.e., Table 6 of Schlegel et al.\ 1998). The star will be 3 mag (15 times) fainter at $I$ in the spectral region where we are observing than the normal stars for which we obtained good data, and be at the faint end ($K\sim 10.1$) in Figure~2.}. After removing these objects, only 3\% of our RSG candidates yielded  nothing but sky. 

\section{Determining LMC Membership}
Armed with spectra of nearly 2000 stars, it was time to determine which ones are actual LMC supergiants. This question is most pertinent for the YSGs because of the large foreground contamination expected. In the following section we describe how we used the stars' radial velocities to weed out the foreground dwarfs from the LMC supergiants.

\subsection{Radial Velocities}
As mentioned earlier, radial velocities are the key to separating foreground dwarfs from YSGs. The LMC has an average radial velocity of 278 km s$^{-1}$ (Richter et al.\ 1987) with a rotational velocity of a few 10s of km s$^{-1}$ superimposed upon that (Kunkel et al.\ 1997, Kim et al.\ 1998). Thus, stars in the LMC will have radial velocities centered around this value, while stars in our own Milky Way will have radial velocities centered around 0 km s$^{-1}$ with a spread that we don't expect to overlap with the velocities of LMC stars. Further details will be discussed in \S4.

To determine the radial velocities, we used the spectra of our observed radial velocity standards. Our wavelength range of $7700 - 9200$ \AA\ included the Ca II triplet ($\lambda \lambda$ 8498, 8543, 8662) which is known to be strong over a large temperature regime (which conveniently includes the temperature regime of YSGs, yellow dwarfs, and RSGs). Before the velocity calculations, we normalized the spectra using a 9th order cubic spline and subtracted 1.0 to remove the continuum. All velocity calculations were done using IRAF's cross-correlation package ``fxcor." Cross-correlating the radial velocity standards against each other using a wavelength range surrounding the Ca II triplet ($8400 - 8700$ \AA) resulted in small uncertainties of $\sim$1 km s$^{-1}$. We were then able to cross-correlate our candidate spectra against the standards using the same wavelength range.

As Neugent et al.\ (2010) found with YSG candidates in the SMC, many of our observed LMC candidates were sufficiently early that the Ca II triplet was contaminated by Paschen Balmer lines (i.e., P16, P15 and P13 at $\lambda \lambda$ 8502, 8545, and 8665, respectively). To rectify this, we instead used the Paschen hydrogen lines from P11 to P19 for cross-correlation of the stars that showed Paschen lines by visual inspection. For templates, we chose a few ``exemplary" Paschen-lined spectra and measured their radial velocities by hand. For cross-correlation of these stars, we used a wavelength range surrounding Paschen lines P11 to P19 ($8400 - 8900$ \AA).

We next used three different methods to understand the errors of our radial velocities: the Tonry and Davis (1979) $r$ parameter, the internal errors from the fits, and the external errors from multiple observations of the same objects.

The $r$ parameter measures how well the cross-correlation worked where larger values indicate more reliable results. For example, the average $r$ parameter for Paschen lined stars after being cross-correlated against our Ca II triplet radial velocity standards was 20.3. However, the average $r$ parameter for these stars after being cross-correlated against Paschen-lined templates was 53.8. Overall, the average $r$ parameter for YSG candidates was 79.4 (comparable to the value of 75.7 found by Neugent et al.\ (2010) for SMC YSG candidates), while the average value for RSG candidates was 44.2. We believe that the average RSG $r$ parameter is lower than the average YSG $r$ parameter because given our color selection criteria, the RSG candidates are dimmer than the YSG candidates at our observed wavelength range. Still, even a $r$ parameter of 44 is impressive (for example, Tonry and Davis (1979) published single digit values when studying galaxies). 

As expected, internal errors (errors due to uncertainties in the cross-correlation fits) were highly correlated with $r$ parameters. Stars with large $r$ parameters ($r > 100$) had internal radial velocity errors ($\sigma$) close to 1.5 km s$^{-1}$ while stars with average $r$ parameters ($50 < r < 100$) had errors around 3 km s$^{-1}$. Stars with low $r$ parameters ($r < 50$) had errors around 5 km s$^{-1}$. 

There were 179 stars observed twice, which allowed us to calculate an external error of 3 km s$^{-1}$ by determining the mean absolute difference between two observations. These errors most likely stem from our ability to fit the comparison lines and not any lack of signal to noise in our spectra. 

\begin{figure}
\epsscale{0.5}
\plotone{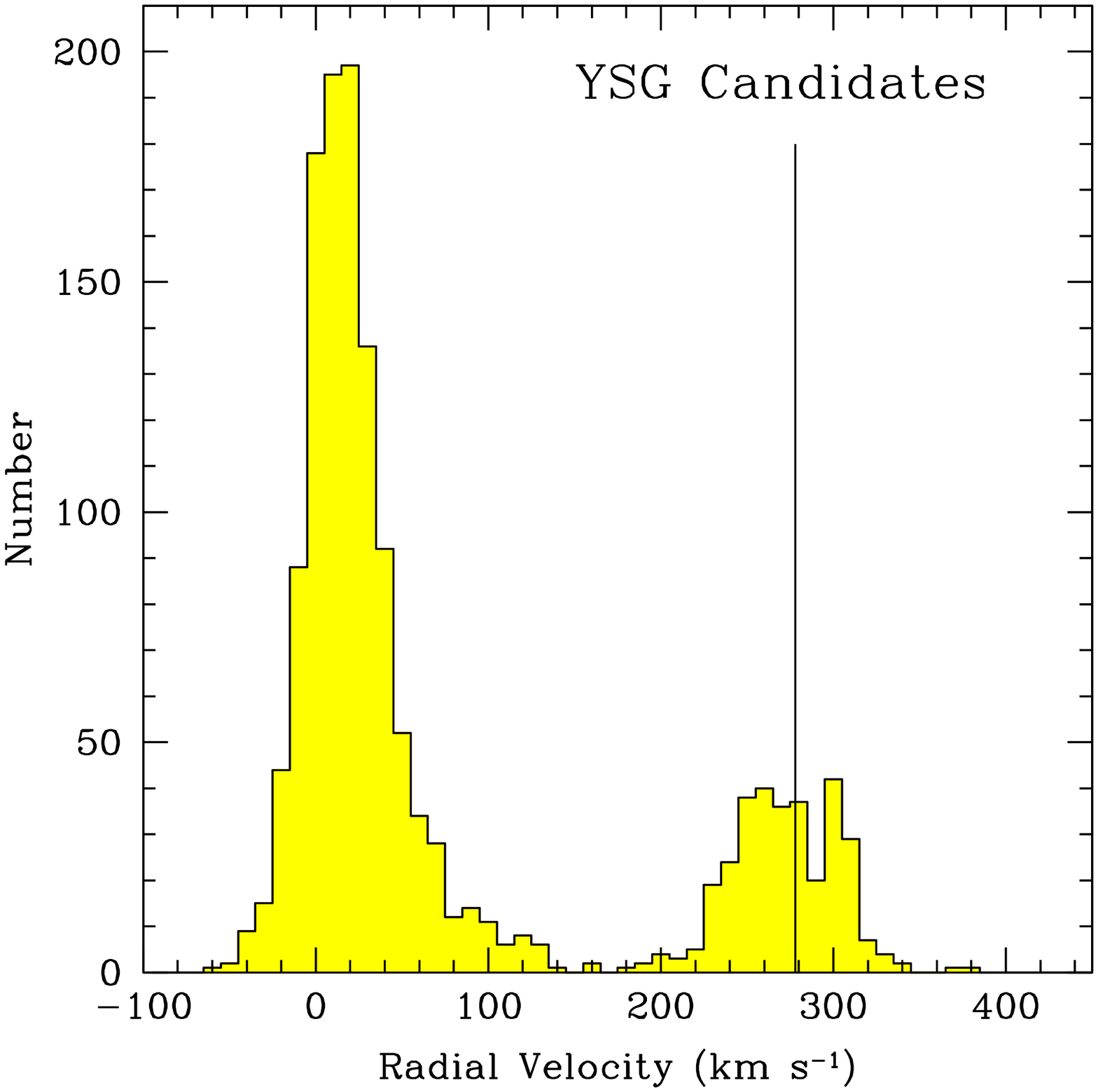}
\epsscale{0.5}
\plotone{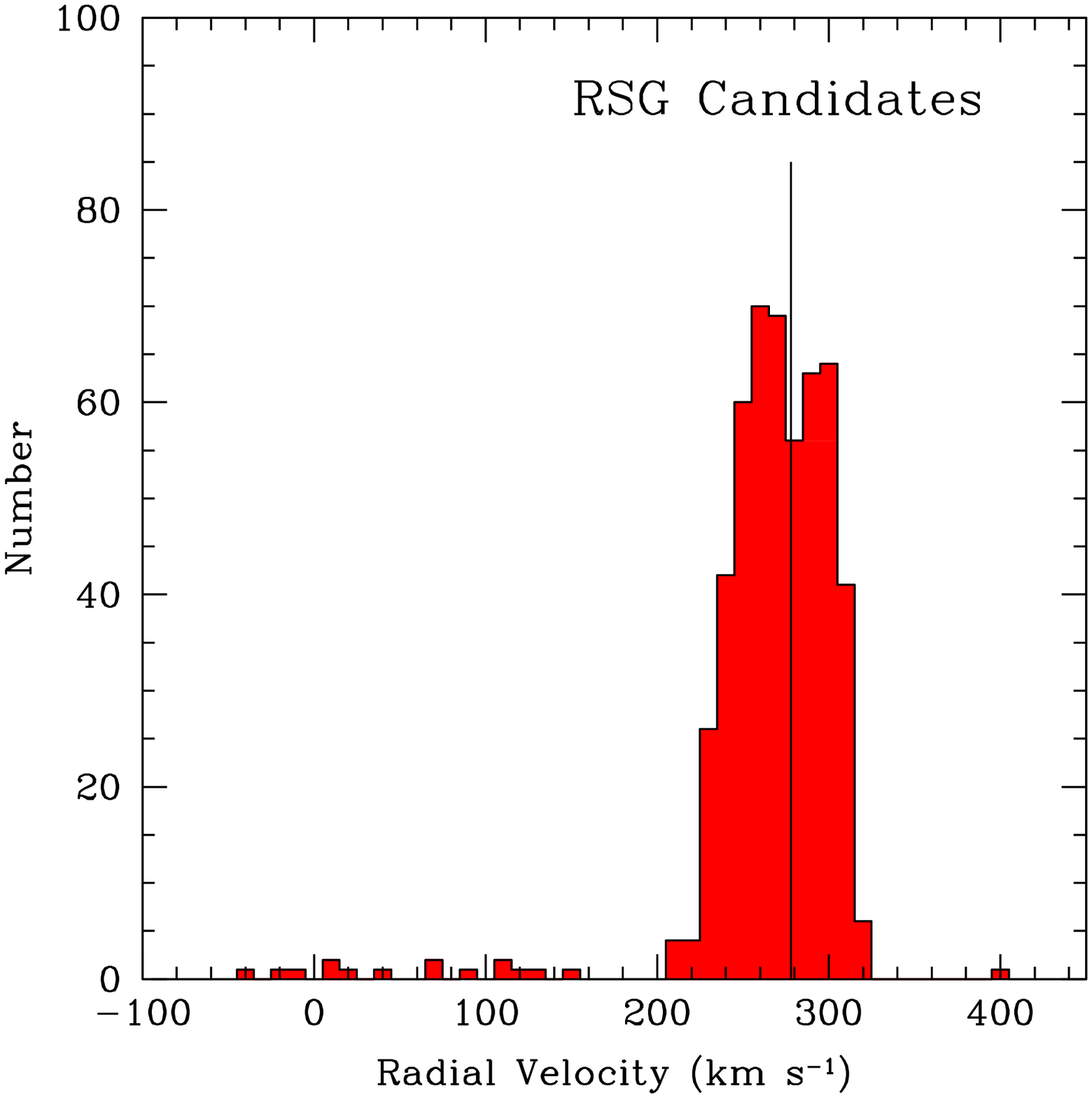}
\caption{\label{fig:velHist} YSG and RSG candidate radial velocity histograms. The bimodal distribution shows the clean separation between the foreground stars (centered around 0 km s$^{-1}$) and the LMC supergiants (centered around 278 km s$^{-1}$ and shown by the vertical line).}
\end{figure}

Identifying LMC supergiants based on radial velocities proved to be quite trivial. As Figure~\ref{fig:velHist} shows, there is a clear bimodal distribution among the stars' radial velocities with foreground dwarfs clustered around 0 km s$^{-1}$ and LMC supergiants clustered around 278 km s$^{-1}$. This figure further confirms that we were able to successfully identify LMC RSGs using color and proper motion criteria alone while radial velocities were necessary for LMC YSG confirmation. Our final radial velocity results, along with other identifying information about each supergiant candidate, are shown in Tables~\ref{tab:allobsYSG} (YSGs) and \ref{tab:allobsRSG} (RSGs). 

In their study of SMC YSGs, Neugent et al.\ (2010) found a small but significant subsample of stars that fell in between the stars that were clearly Milky Way or clearly SMC members. For the hotter of these, they used the OI $\lambda 7774$ line to help assign membership. Osmer (1972) had shown the OI $\lambda 7774$ line is sensitive to luminosity in Galactic F-type stars, a dependence that Przybilla et al.\ (2000) showed was due to non-LTE and sphericity effects. Drout et al.\ (2009) showed that this worked for even G-type stars in M31, a high metallicity environment. At the low metallicity of the SMC, Neugent et al.\ (2010) did not find as good a correlation between the OI $\lambda 7774$ line strength and luminosity for the cooler ($T_{\rm eff}< 5200$) stars in their sample. Fortunately, as we've seen previously, the radial velocities by themselves did a more than adequate job here thanks to the considerably larger systemic velocity of the LMC compared to the SMC (278 km s$^{-1}$ versus 158 km s$^{-1}$). The use of the OI $\lambda 7774$ will be discussed further in a future paper.

\subsection{Final Membership Assignments}
\begin{figure}
\epsscale{1}
\plotone{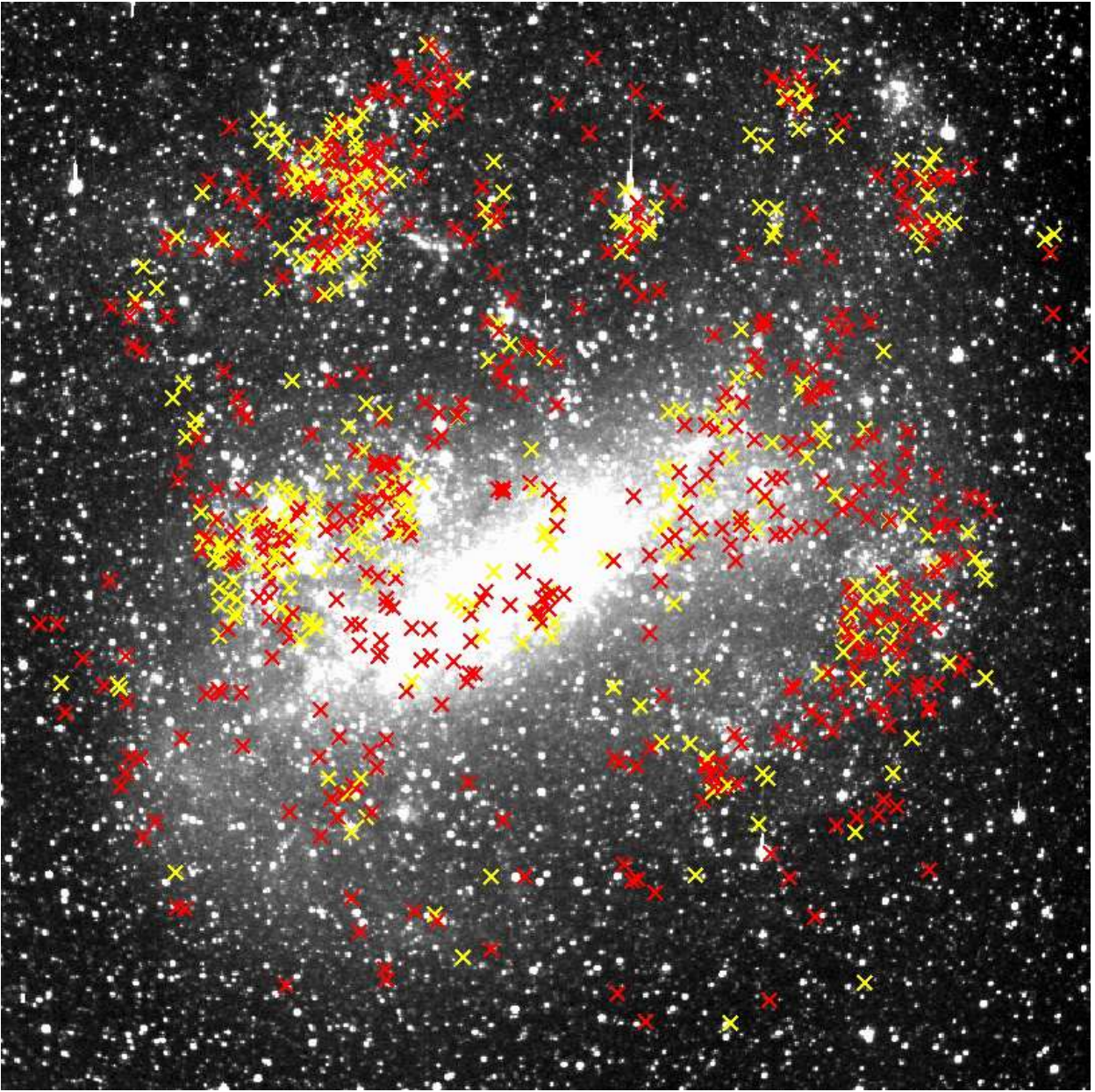}
\caption{\label{fig:configCat12} Locations of confirmed yellow and red supergiants. The yellow $\times$s represent our confirmed YSGs and the red $\times$s represent our confirmed RSGs. Note the good spatial agreement between the YSGs and RSGs; compare to Figure~\ref{fig:config}.}
\end{figure}

Owing to the large radial velocity separation between the Milky Way and the LMC, membership determination proved to be straight forward. Based on a visual examination of Figure~\ref{fig:velHist}, we concluded that stars with radial velocities higher than 200 km s$^{-1}$ (317 YSG candidates and 505 RSG candidates) are probable LMC supergiants and were labeled ``category 1." Stars with radial velocities lower than 155 km s$^{-1}$ (1129 YSG candidates and 17 RSG candidates) are probable foreground dwarfs and were labeled ``category 3." The 6 candidates (all YSG candidates) between these two radial velocity cut-offs were then labeled ``category 2" or possible, but not probable, LMC supergiants. The locations of the category 1 and category 2 YSGs and RSGs within the LMC are shown in Figure~\ref{fig:configCat12}. Note the excellent spatial agreement between the YSGs and RSGs. Compare this to Figure~\ref{fig:config}, where the YSG {\it candidates} were heavily contaminated by foreground stars, and thus show a far more uniform spatial distribution.

A subset of our RSG candidates had previously been observed by Massey \& Olsen (2003) and confirmed as LMC RSGs using radial velocities. These 43 stars are indicated in Table~\ref{tab:allobsRSG}. A comparison of the radial velocities presented in this work versus the radial velocities presented by Massey \& Olsen (2003) yields a average difference of 3.6 km s$^{-1}$ with a standard deviation of 4.3 km s$^{-1}$, consistent with our own external radial velocity errors.

\section{Contamination and Completeness}
With LMC membership determined, we focused on understanding our contamination and completeness rates for both YSGs and RSGs.

\subsection{Foreground Dwarfs}
Before the observing proposal was even submitted, we estimated the foreground contamination for both YSGs and RSGs using control fields. The vast majority of these contaminants should be Milky Way disk dwarfs, which we knew we could separate from the LMC's population of supergiants using radial velocities. Stars in these control fields were selected using the same criteria described in \S\ref{sel} and were positioned at $\pm 5$ degrees in Galactic longitude from the LMC's center. Since these fields should be populated solely by foreground stars, they provide a direct contamination estimate. 

As described in \S\ref{sel}, we began with 2187 YSG candidates. The two control fields yielded 1471 and 1751 stars, respectively. Thus, we estimated our foreground dwarf contamination to be between 67\% and 80\%. After observing these candidates, we found a contamination of 78\%, as expected. 

We began this project quite confident that we could estimate the number of RSGs in the LMC based on color selection criteria and proper motion cut-offs alone, as discussed in \S\ref{intro}. Thus, we expected the foreground dwarf contamination to be quite minimal. Out of the 522 observed RSG candidates, 505 turn out to be RSGs. This yields a contamination rate of 3\%, as expected.

\subsection{Halo Giants}
While foreground dwarfs can be eliminated using their radial velocities, this is not the case with halo giants as their radial velocities may overlap with that of the LMC, as much of the LMC's systemic velocity is actually the reflex motion of the sun (Courteau \& van den Bergh 1999). We can estimate their relatively minimal contamination using the Besan\c{c}on models (Robin et al.\ 2003). Figure~\ref{fig:besAll} shows radial velocity histograms based on the model results for stars within the same color range as our YSG and RSG candidates.

Since we observed 70\% of our YSG candidates, the models predict that 3 (50\%) of our category 2 YSGs and 8 (3\%) of our category 1 YSGs are halo giants. This suggests that while several of our category 2 yellow stars may actually be Milky Way members, it is very unlikely that any of our category 1 YSGs are Milky Way halo giants.

Similarly, for the red stars, the Besan\c{c}on models predict only one red halo giant with a radial velocity larger than 155 km s$^{-1}$. Thus, since we observed 44\% of our RSG candidates, there is a low probability that our RSG sample is uncontaminated by more than one halo giant.

\subsection{Known Supergiants Not Observed}
To gain a greater understanding of our survey's completeness, we conducted a literature search of previously known YSGs within the LMC that our survey did not find. The results are shown in Table~\ref{tab:knownYSG}\footnote{Evans et al.\ (2011) identify two LMC members of type G5/K3 that are not on our list, VFTS 289 and [P93] 2186. These are not included in Table~\ref{tab:knownYSG} as they are likely cooler than the stars we consider here.}. Of the eleven ``known" LMC YSGs, nine were on our original observing list but weren't observed and the remaining two never made it onto our list for the reasons explained in the table. While our original target list may not have included every LMC yellow supergiant, this literature reality check suggests that we're only missing a few, rather than tens of stars.

Previous surveys for RSG members of the LMC are largely incomplete; see Massey \& Olsen (2003).  Objective prism surveys, such as the case study by Sanduleak \& Phillip (1977), were used by Humphreys (1979),  Elias et al.\ (1985), and Oestreicher \& Schmidt-Kaler (1998) for follow-up studies, but the poor precision of the coordinates render them of limited use for all but the brightest stars. Massey \& Olsen (2003) and Levesque et al.\ (2006) used the {\it UBVRI} survey of the LMC by Massey (2002) for follow-up RSG work, but this survey only covered part of the LMC. Massey \& Olsen (2003) used radial velocities to confirm that the foreground contamination of RSGs was less than $\sim$10\%, while here we find that the combination of 2MASS photometry and UCAC-3 proper motions reduces the contamination to $<3\%$. Regardless, given that the contamination is on the order of 10\%, we can certainly estimate the number of RSGs in the LMC to a reasonably accurate value, about 1800, for $K<10.2$.  Cross-correlating this list against the literature (especially given the poor coordinates for the objective prism studies) is beyond the scope of the present paper.

\section{Testing the Current Geneva Evolutionary Models}
Here we describe how we used the stars' colors to determine their temperatures and luminosities and how these results allow us to comment on the accuracy of the current Geneva evolutionary models.

\subsection{Determining Temperatures and Luminosities}
\label{trans}
To place the stars on the H-R diagram, we need to determine the stars' effective temperatures and bolometric luminosities. For the YSGs, Neugent et al.\ (2010) demonstrated that $B-V$ gave slightly better results than (say) $J-K$. However, not all of the stars in our sample have good $B-V$ colors, but all do have good $J-K$ colors from 2MASS since this was part of our selection process. In addition, we would like to determine physical properties for the RSGs. For these, $V-R$ is preferable (see discussion in Drout et al.\ 2009) but again, we have $V-R$ colors only for a limited subset of our sample. For consistency, we used $J-K$ for the effective temperature determinations of all our stars, but checked for systematic issues using the other colors as follows.

The transformations for the YSGs are newly derived here, using the Kurucz (1992) ATLAS 9 atmospheres. For the RSGs, we adopt the transformations determined using the MARCS models as described by Levesque et al.\ (2006).  For all stars, we first converted the 2MASS $J-K$ colors to the ``Bessell \& Brett (1988) Homogenized System" following Carpenter (2001): $$J-K = \frac{(J-K)_{\rm{2MASS}} + 0.011}{0.972}$$ We next de-reddened the photometry assuming that $E(J-K)=0.535 \times E(B-V)$ (Schlegel et al.\ 1998), and adopted a constant value $E(B-V)=0.13$ (Massey et al.\ 1995); i.e., $E(J-K)\sim 0.07$. Note that while this value is appropriate for early-type stars, the reddening is probably greater for RSGs; see discussion in Massey et al.\ (2005) and Levesque et al.\ (2005, 2006)\footnote{This is another advantage of using $J-K$, as the answers we get are less sensitive to the assumptions of constant reddening.}.

When $-0.07<(J-K)_0\leq0.70$:
\begin{eqnarray}
\label{Equ-hot}
\log T_{\rm eff}=3.968-1.347 (J-K)_0+5.173 (J-K)_0^2-14.665  (J-K)_0^3+\\ \nonumber 22.098  (J-K)_0^4-16.277  (J-K)_0^5+4.633 (J-K)_0^6
\end{eqnarray}

For cooler stars, the transformation obtained using the MARCS models is simple; note that there is no logarithm involved. 

When $1.4\ge (J-K)_0>0.70$: 
\begin{equation}
\label{Equ-cool}
T_{\rm eff}=5638.0-1746.2 (J-K)_0
\end{equation}
Care has been taken to avoid a discontinuity; at $(J-K)_0=0.7$ Equation~\ref{Equ-hot} and Equation~\ref{Equ-cool} each yield a value of $\sim$4415~K.

We derived similar transformations for $B-V$ for the yellow stars\footnote{For stars with $-0.08<(B-V)_0<1.78$, $\log T_{\rm eff}= 3.923-0.755 (B-V)_0+2.2065 (B-V)_0^2-3.9777 (B-V)_0^3+3.7084 (B-V)_0^4 -1.7070 (B-V)_0^5+0.3053 (B-V)^6$.} and adopted the Levesque (2006) transformations from $V-R$ for the cooler stars\footnote{$T_{\rm eff}=7798.3-7824.4(V-R)_0+4554.8 (V-R)_0^2-905.21 \times (V-R)_0^3.$}. For the former, the comparison with the results for $J-K$ (Equation~\ref{Equ-hot}) yields an average difference of $-0.015$~dex in log effective temperature, or 175 K at 5000 K, an insignificant difference.  (The standard deviation for the sample of 206 stars is 0.04~dex.)  For the cooler
stars, the comparison with the $J-K$ (Equation~\ref{Equ-cool}) is $-0.003$~dex for the sample of 295 stars, with a standard deviation of 0.02~dex. Thus, we do not feel the analysis is compromised by the lack of optical photometry for these stars.

The bolometric luminosity was computed using the $K$ value transformed from 2MASS again following Carpenter (2001): $K=K_{\rm 2MASS}+0.044$. It was then corrected for interstellar reddening by 0.05~mag, i.e., $K_0=K-0.367 \times E(B-V)$, following Schlegel et al.\ 1998.  The bolometric correction at $K$ (BC$_K$) is positive, and transformations were computed using the ATLAS9 and MARCS colors as above. 

For $3.60<\log T_{\rm eff}<4.04$, 
\begin{equation}
\label{Equ-BChot}
{\rm BC}_K = 28.48-7.244 \times \log T_{\rm eff}
\end{equation}

For cooler stars, the relation is nearly linear with effective temperature, and not the logarithm of the temperature. Thus for $T_{\rm eff}<4000$ K,
\begin{equation}
\label{Equ-BCcool}
{\rm BC}_K=5.50-0.739 \times \frac{T_{\rm eff}}{1000}, 
\end{equation}
following Levesque et al.\ (2006). Again, care was taken to minimize any discontinuity, and both equations give a BC of 2.4-2.5 at $\log T_{\rm eff}=3.60$, or $T_{\rm eff}$=4000 K.

The computed effective temperatures and luminosities are shown for our category 1 and 2 LMC supergiants in Table~\ref{tab:derived}.  The typical uncertainty in the 2MASS photometry is 0.02~mag in $K$ and 0.03 in $J-K$.  This
propagates to errors of 0.005~dex in $\log T_{\rm eff}$ and 0.05~dex in $\log L/L_\odot$ for the RSGs,
and  0.015~dex in $\log T_{\rm eff}$ and 0.10~dex in  $\log L/L_\odot$ for the YSGs.

\subsection{Stars That Are Too Red}
\label{Sec-toored}
The previously discussed transformations are determined only for $(J-K)_0\le 1.4$, corresponding to a MARCS model temperature of 3000 K, since the coolest RSGs identified in the LMC have effective temperatures of 3450 K (Levesque et al.\ 2006), corresponding to $(J-K)_0\approx$1.25, or a $(J-K)_{\rm 2MASS} \approx$1.27. This 3450 K limit is consistent with the Hayashi limit, set by the demands of hydrostatic equilibrium (Hayashi \& Hoshi 1961). The coolest of these are unusual variables (Levesque et al.\ 2007), exhibiting non-periodic large swings in effective temperature on the order of a year. The only known stars that are redder have high circumstellar extinction, such as the heavily enshrouded and unusually cool RSG WOH G64 (Levesque et al.\ 2009).

In Figure~\ref{fig:JMKvsK}, we see there is a sharp break near $(J-K)_{\rm 2MASS} \approx$1.25. Yet, there are stars extending redwards beyond $(J-K)_{\rm 2MASS} \approx$2. When observing, we found occasional instances of blank sky where we expected legitimate sources; invariably these objects were in the ``too red" region, as discussed in \S2.3. For others, we have valid radial velocities that indicate LMC membership.  

To understand the nature of these objects, we carefully checked VizieR for all of our objects with $(J-K)_{\rm 2MASS}>1.2$, and list what we found in Table~\ref{tab:toored}. Many of these stars have some emission near the 10$\mu$m Si peak, as indicated by a detection in the IRAS, MSX, or AKARI surveys (Moshir et al.\ 1989, Egan et al.\ 2003, and Ishihara et al.\ 2010, respectfully), indicating significant dust production. This would of course explain the much too red colors.

Most of the stars have an entry in the OGLE database (Soszy\'{n}ski et al.\ 2009). However, many of the periods appear to be incorrect since the light curves look very poor. We suspect that many of the nominal periods are just aliasing or noise in the data-taking cadence (Stothers \& Leung 1971). Some of the stars are also listed as Mira variables by Soszy\'{n}ski et al.\ (2009). This is not impossible; although our survey is aimed at detected massive red supergiants, it is certainly true that AGBs and Miras could contaminate the faint end of our sample, even though we strove to eliminate these stars by not observing lower mass supergiants, as discussed in \S2.1\footnote{Indeed, there are a few stars in the list that have been classified by Lundgren (1988) as having enhanced ZrO (``S-type", i.e., M4S), indicative of their being AGB stars on their way from evolving from M-type to carbon-type.}. But, amongst all the OGLE light curves (many more were examined than those listed in Table 5) we saw nothing that resembled an ordinary Mira. Typically Miras have periods longer than $\sim$150 days and large amplitudes, $\sim$2 mags or more. By contrast, our observed overly red stars luminous enough to be in the ASAS database are characterized by two sorts of light curves. There are the types whose light curves are called ``Lc" in the GCVS (Samus et al.\ 2006) which have slow, irregular variations with cycle lengths of a few hundred days (like a typical RSG), and then there are the others which have much smaller, short-term variations (of order 100 days) and very long (thousands of days) cycles superposed on that of half a mag (or more) amplitudes. We have flagged these in Table~\ref{tab:toored}.

We include in Table~\ref{tab:toored} the period in days; these are all from the OGLE catalogue except for two from the GCVS. We confirmed both of the latter using the ASAS-3 database (Pojmanski 1997). The OGLE periods are quoted only if the light curves are reasonably smooth.

In a few cases, the OGLE phased plot shows a very long cycle ($\sim$1200 days) which additionally contains a much shorter period or periods. These stars are most likely semi-regulars that have very long overtone periods of 10-20$\times$ the fundamental period. In other cases we give notes indicating that the star is variable in an irregular manner, like one expects for the luminous M supergiants, or that it has a very long variation in the ASAS-3 series.

For the stars redder than the reddest RSGs, we would have underestimated the effective temperatures and bolometric luminosities\footnote{Note that not only are the bolometric corrections to the $K$-band positive, but they also increase with decreasing effective temperature in this temperature regime.} by simply extrapolating Equations~\ref{Equ-cool} and \ref{Equ-BCcool} for larger $(J-K)_0$ values. This is particularly an issue for stars with 10$\mu$m emission, and hence possibly substantial amounts of dust. Therefore, we did not include the 33 stars with $(J-K)_0\ge 1.3$ in the HRD. We also refrain from listing their (extrapolated) physical properties in Table~\ref{tab:derived}, as we expect the values would be misleadingly cool and under-luminous. These stars are worthy of spectroscopic follow-up, and we are beginning such efforts.

\subsection{The H-R Diagram}
We ultimately aim to compare the relative numbers of different kinds of massive stars (O-type main-sequence, and He-burning YSGs, RSGs, and WRs) as a function of mass throughout the Local Group galaxies forming stars. This will provide us with the observational database against which current and future models of stellar evolution models may be compared. Currently, data are too incomplete to accomplish this (see discussion in Massey 2010), but interesting tests can be conducted with what we have at hand. The two tests we make here are: (1) Do the evolutionary tracks extend to the appropriate effective temperatures and luminosities for the YSGs and RSGs? (2) Are the relative lifetimes predicted by the models as a function of luminosity correct? We can answer this second question only for the YSGs in our sample, as stars of the same mass overlap considerably in luminosity for the RSGs, as the evolutionary tracks become nearly vertical.

\begin{figure}
\epsscale{0.9}
\plotone{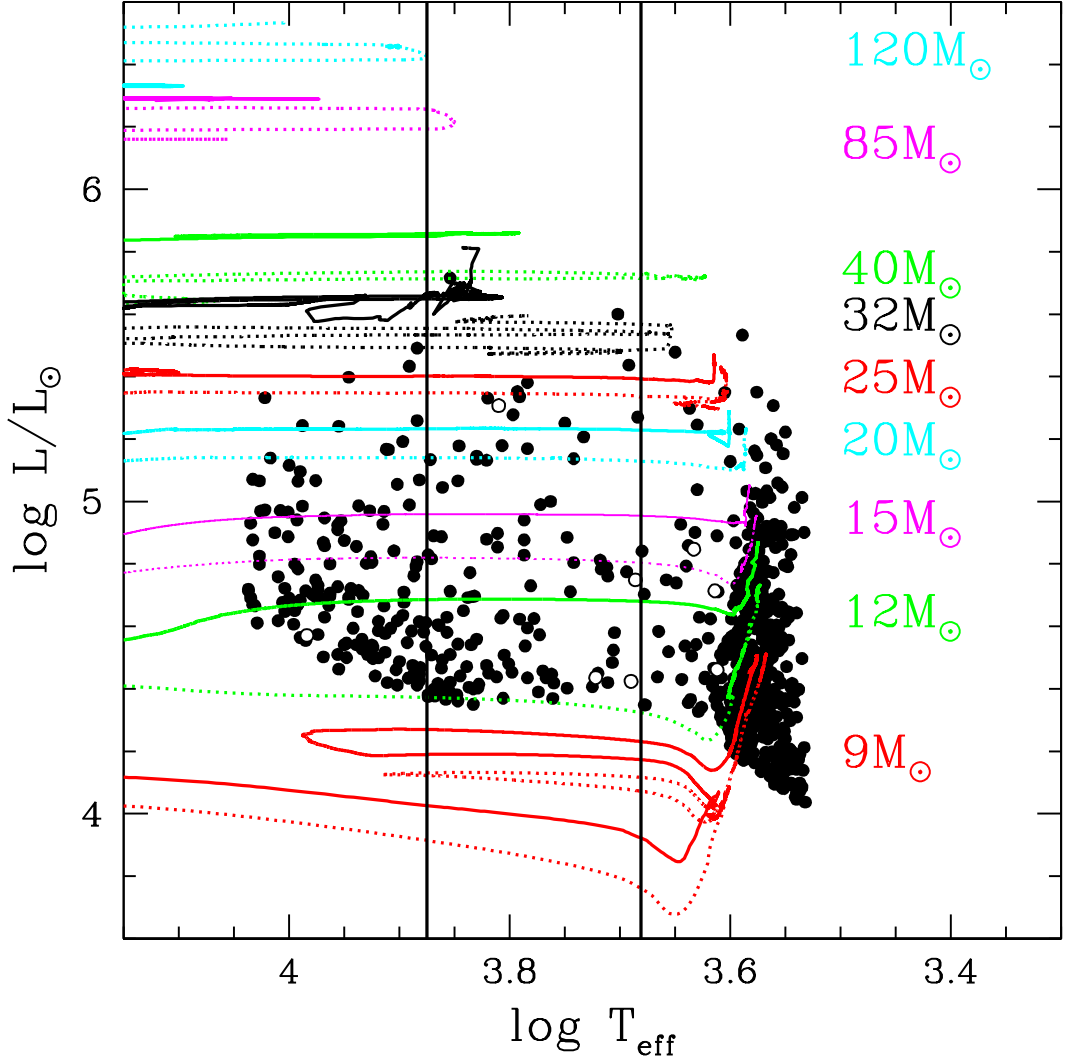}
\caption{\label{fig:HRD}The H-R Diagram. The probable LMC members (category 1) are showed by solid points; the less certain members (category 2) by open circles. The typical errors are comparable to the point size.
The $z=0.006$ Geneva evolutionary tracks (Chominne et al.\ 2012, in prep) are shown. The solid curves denote the models run with an initial rotation velocity 40\% of the critical value, while the dashed curves denote the models with no initial rotation. The initial masses are indicated on the side. The two vertical lines delineate the YSG region while stars at the far right are the red supergiants. }
\end{figure}

The HRD is shown in Figure~\ref{fig:HRD}. The latest Geneva $z=0.006$ evolutionary tracks, which follow massive star evolution all the way to the end of the core carbon burning phase (Chomienne et al.\ 2012, in prep), are shown in color, with the initial masses indicated\footnote{Identical Geneva evolutionary models at a higher metallicity ($z=0.014$) have been published with full details given by Ekstr\"{o}m et al.\ (2012). The newer models discussed here were simply computed at a lower metallicity, with the appropriate scaling of mass-loss rates.}. The solid curves denote the tracks computed assuming an initial rotation speed of 40\% the critical (breakup) speed. The 40\% value is consistent with zero-age solar-metallicity B-type stars; see discussion in \S3 of Ekstr\"{o}m et al.\ (2012). This value may be too high or too low for other mass ranges, but is suitable for the 12-30$M_\odot$ range. Individual stars may have slower or faster rotation thus, these tracks represent an average behavior. The tracks computed with no rotation are shown by dashed lines for comparison. Note that all tracks were computed using the Asplund et al.\ (2009) lower solar abundance values, and thus the $z$ value for the LMC is lower than what we would have previously used ($z=0.008$). 

We show the location of our stars based upon the transformations described in \S\ref{trans} in Figure~\ref{fig:HRD}. Stars that are probable LMC members (category 1) are shown with solid points while the six less certain (category 2) stars are shown as open circles. As discussed in \S\ref{trans}, the uncertainties in the placement of these points are small, comparable to the size of the points.  At the low luminosity end, the distribution has been ended by the magnitude cutoff used when selecting our sample, which will extend to lower luminosities at cooler temperatures. The YSG realm is delineated by the two black vertical lines denoting the effective temperature range of 4800 - 7500 K. Stars to the left of this are blue supergiants, and our selection criteria was overly generous by including them to 10,000 K ($\log T_{\rm eff} =4.0$). Stars to the right are red supergiants, and they pile up at the Hayashi limit, set by the coolest temperature and largest radius (for a given mass) for which a star can be in hydrostatic equilibrium (Hayashi \& Hoshi 1961; see recent discussion in Levesque et al.\ 2007). This limit is a function of the metallicity, with the tracks shifting about 0.03~dex to higher temperatures as the metallicity lowers from solar to that of the LMC (see Figure 10 in Levesque et al.\ 2006 and Figure 1 of Drout et al.\ 2012).

The 32$M_\odot$ track shown (in black) represent a kind of transition track.  Stars with lower masses should remain in the RSG region, while those with higher masses evolve back to the blue side, becoming Wolf-Rayet stars. The loops are indicative of some unstable situation reflecting the fact that the model is ``hesitating" between the red and the blue.  Such unstable behavior is in fact known for some high luminosity YSGs, as discussed in \S\ref{Sum}.  Note that the exact mass for such a transitional situation is dependent upon the assumed mass-loss rate during the RSG phase, which is still uncertain.

The agreement between the locations of the stars and evolutionary tracks in the HRD and the evolutionary tracks is nothing short of exceptional. First, we find that the tracks correctly reproduce the locus of the RSGs, not only in terms of the effective temperatures, but also in terms of the upper luminosity limit. Most RSGs have masses of 15$M_\odot$, but a few are found at higher masses.  The highest luminosity RSGs we find have $\log L/L_\odot \sim 5.5$, and this is also the limit where the tracks no longer extend to the RSG stage.  We note that the 20 and the 25$M_\odot$ tracks  could extend to slightly cooler temperatures, but the agreement is otherwise excellent. The rotating 32$M_\odot$ model predicts no $\log L/L_\odot =5.6$ RSGs should be seen, and indeed none are. The presence of two stars at cooler temperatures and slightly lower luminosities  than the rotating 32$M_\odot$ track are consistent with our expectation that tracks of slightly lower mass than the 32$M_\odot$ would reach to cooler temperatures.  

The locations of the YSGs are similarly consistent with the tracks. The highest luminosity in our sample is found at $\log T_{\rm eff}\sim 3.9$ and $\log L/L_\odot \sim 5.7$, a good match to the 32$M_\odot$.  At this luminosity we don't find any cooler YSGs, and indeed the 32$M_\odot$ track does not extend to cooler temperatures. The higher luminosity tracks (120$M_\odot$ and $85M_\odot$) do not extend into the YSG regime, nor do we find any stars at the corresponding high luminosities.  The presence of a single cooler YSG at $\log L/L_\odot=5.6$ is consistent with interpolating between the 25 and 32$M_\odot$ (rotating) tracks, again recalling that the 32$M_\odot$ is at a transition between subsequent blue-wards evolution and not. 

We can now test the models quantitatively, using the lifetimes predicted by the models. As long as our sample is unbiased in luminosity, then the {\it relative} number of YSGs as a function of luminosity (mass) should scale as the lifetimes of the YSG phase with a small correction for the initial mass function: $$N_{m_1}^{m_2} \propto [m^\Gamma]_{m_1}^{m_2} \times \bar \tau$$ where $\Gamma$ is the slope of the initial mass function (taken here to be $-1.35$ following Salpeter 1955; see also Massey 1998), and $\bar \tau$ is the average duration of the evolutionary phase for stars with masses between $m_1$ and $m_2$.    

We list the predicted lifetimes of the YSG phase in Table~\ref{tab:lifetimes}. Note that the times are given in terms of {\it thousands} of years; the YSG phase is indeed short-lived! We give the comparison between the model predictions and the observed number in Table~\ref{tab:ratios}, normalized to the 12-15$M_\odot$ bin.

The relative lifetimes predicted by these newer models show excellent agreement with the observations, unlike what we found in M31 and the SMC (Drout et al.\ 2009, Neugent et al.\ 2010) using older versions of the Geneva models\footnote{The differences between the older Geneva models and the newer ones presented here are described in Ekstr\"{o}m et al.\ (2012). They can be summarized as follows: (1) The initial composition differs. In the newer models a mixture of heavy elements based on the solar abundances obtained by Asplund et al.\ (2005) is used. (2) The opacity tables differ due to the changed initial composition. (3) The nuclear reaction rates have been updated. (4) A different prescription for the mass loss rate during the RSG stage has been adopted. (5) A different prescription for the diffusion coefficient describing rotational mixing has been adopted.}.  The agreement with the ``S4" (initial rotation 40\% of the critical breakup speed) shows astonishingly good agreement.  The physically unrealistic case of no rotation (``S0") shows poorer agreement. The only area where the models may predict an overabundance of YSGs is for the 32$M_\odot$ track, with its significantly longer lifetime, due to the many loops.  Even so, the agreement for it is much better than what was found for M31 (Drout et al.\ 2009) and the SMC (Neugent et al.\ 2010).

In making the comparisons between the expected relative number of YSGs and the expected number of YSGs we have averaged the lifetimes of the models in order to compute the lifetime in a particular mass bin.  However, we expect that the lifetime of the $32M_\odot$ track is not representative of models of slightly lower or higher mass, since the $32M_\odot$ is a transitionary model.  Thus we also include in Table~\ref{tab:ratios} the comparisons we obtain by ignoring the $32M_\odot$ track.  When we do so, we find that the agreement between observation and the models for the YSGs is excellent.

Note that the Geneva models follow the evolution to the end of core carbon burning. Thus, unless dramatic mass loss occurs at the very end, the tracks are complete as shown. If high mass loss occurs does occur at the very end, the star may indeed become blue again, crossing from the RSG position into the YSG region. However, the duration will be so short that it will not change the results. 

An assumption in our calculation is that, averaged over the LMC as a whole, star formation has stayed essentially constant over the past 20~Myr, since the models predict that the highest mass YSGs came from stars that formed 3-5~Myr ago, while the lowest mass YSGs came from stars that formed 20~Myr ago. Harris \& Zaritsky (2009) have investigated the star formation history of the LMC, and conclude that over the past 5 Gyr the star formation rate has stayed constant to within a factor of 2; in recent times, there was a ``mini-burst" $\sim$12~Myr ago, but the present star formation rates are comparable to this today.  This burst would correspond to the main sequence lifetime of a 15-20$M_\odot$ star according to the models.

We can perform one other reality check. The {\it relative} lifetimes of the models appear to be solid for the YSG phase, but what about the absolute lifetimes? Drout et al.\ (2009) argue from an (admittedly highly uncertain) number of O stars in M31 that the average lifetime of the YSG phase should be $\sim$3,000 years, much shorter than what the models predict. Let us consider the case here for the LMC. Massey (2010) estimates there are $\sim$6,000 main-sequence stars in the LMC with masses $>20 M_\odot$. This number is probably no worse than a factor of 2 or 3. The average IMF-weighted main-sequence lifetime, according to the S4 models, is $\sim$6.5 Myr.  We observed 11 YSGs with masses above 20$M_\odot$ (Table~\ref{tab:ratios}), and we observed $\sim$70\% of the candidates. Thus we might expect there to be $\sim$16 high mass YSGs in total. If true, then the average lifetime should be $6.5$ Myr $\times \frac{16}{6000}$, or $\sim$17 thousand years. This is a bit smaller than the ages listed in Table~\ref{tab:lifetimes} for masses above 20$M_\odot$, but we are well within the uncertainties of the number of O stars, in our opinion.

As we emphasized in earlier papers (e.g., Drout et al.\ 2009, Neugent et al.\ 2010), the numbers and location of YSGs are very sensitive to how convection and other mixing processes are treated, as well as uncertain mass-loss rates (Maeder \& Meynet 2000), who quote Kippenhahn \& Weigert (1990) that, ``[The yellow supergiant] phase is a sort of magnifying glass, revealing relentlessly the faults of calculations of earlier phases." In this case it appears that the models may be in great shape.

\section{Summary and Future Work}
\label{Sum}
After observing 1452 potential LMC YSG candidates (70\% of those originally selected), we identified 317 category 1 probable YSGs and 6 category 2 possible YSGs. Similarly, after observing 522 potential RSG candidates (44\% of those originally selected), we identified 505 category 1 probable RSGs.

Confident of our completion, we then placed the stars on the HRD. The new $z=0.006$ Geneva models of Chomienne et al.\ (2012, in prep) do an exemplary job of predicting the relative numbers and locations of different mass YSGs. In a complementary study in M33, Drout et al.\ (2012) found similarly good agreement. However, we have yet to test these models in M31 and the SMC where older models failed since new models are not yet available for such metallicities. But, when they are it will be of great interest to see if the problem with the YSG lifetimes has disappeared. 

Besides allowing us to test the Geneva evolutionary models, this complete sample of LMC YSGs will be useful for many other purposes in the future. Some examples include comparisons with evolutionary tracks, lifetimes, and surface compositions. Additionally, this sample is particularly interesting because it appears YSGs could be core collapse supernova progenitors (see, for example, Maund et al.\ 2011 and Georgy 2012).

Another topic worthy of future study is that of variability. The stars we discuss here are typically more luminous than the classical Cepheids. Only one of the stars in our sample is known to be a Cepheid, J04542376-7054057 (aka as Sk $-70^\circ$ 14, HV 873, and Radcliffe 60), first described by Feast et al.\ (1960), and the derived luminosity ($\log L/L_\odot\sim 4.407$) listed in Table~\ref{tab:derived} is at the extreme lower end of our sample. High luminosity yellow supergiants in the Milky Way are not known for Cepheid-like behavior, although variability can be spectacular. For instance, $\rho$ Cas undergoes strong changes in effective temperature and absolute visual magnitude every 50 years or so (Lobel et al.\ 2003). Luminous Blue Variables (LBVs) may occasionally masquerade as F supergiants, developing a ``pseudo photosphere", as S Doradus did in 2000 (Massey 2000), but the connection between the LBV phenomenon remains speculative (Smith et al.\ 2004 and references therein). Thus, a variability study of a large population of YSGs would be of interest\footnote{Some authors refer to the luminous yellow supergiants as ``hypergiants", a term we eschew as unnecessary hyperbole.}.

The yellow and red supergiants we've identified in the LMC represent a very young ($<$ 20 Myr) population, and our study has produced very good radial velocities. These radial velocities could be used to further study the kinematics of the LMC, following Olsen \& Massey (2007). Their study compared the radial velocities of RSGs to that of carbon stars and the HI gas to argue that we see tidal heating of the stellar disk. It would be of interest to expand this now using the larger data set we have produced here. Similarly, our results suggest a path to further studies of the SMC's kinematics. Neugent et al.\ (2010) used their radial velocity data on the SMC YSGs to briefly comment on the complexity of the kinematics of that galaxy. Now that we have shown that a large sample of RSGs can be identified from the UCAC3, radial velocity studies of such a sample could provide an interesting complement to that of HI studies of the SMC's kinematics.

Overall, we hope to provide a solid observational database against which evolutionary models may be compared. At this point, the LMC's massive star population has been characterized for YSGs, RSGs and WRs (Massey et al.\ 2003). We are currently studying the unevolved LMC OB stars and once this is done, the LMC's massive star population will have been characterized from one side of the HRD to the other.

\acknowledgements
We would like to acknowledge the generous allocation of observing time by the NOAO Time Allocation Committee, and the excellent support we received while observing with Hydra on the Blanco, particularly by Ricardo Venegas. Additionally we'd like to thank Nick Suntzeff for posing a question that improved our discussion. Finally, we would like to thank the anonymous referee who made comments and suggestions which improved the paper. This publication makes use of data products from the Two Micron All Sky Survey, which is a joint project of the University of Massachusetts and the Infrared Processing and Analysis Center/California Institute of Technology, funded by the National Aeronautics and Space Administration and the National Science Foundation.  It also made use of the VizieR catalogue access tool, CDS, Strasbourg, France. This work was supported by the National Science Foundation through AST-1008020.
Facilities: \facility{Blanco}

\begin{deluxetable}{c c c r r r r r r r r c c c l}
\tabletypesize{\tiny}
\rotate
\tablecaption{\label{tab:allobsYSG} Properties of Observed YSG Candidates\tablenotemark{*}}
\tablewidth{0pt}
\tablehead{
\colhead{2MASS}
&\colhead{$\alpha$(J2000)}
&\colhead{$\delta$(J2000)}
&\colhead{[M2002]}
&\colhead{$K_{\rm 2MASS}$}
&\colhead{$J-K_{\rm 2MASS}$}
&\colhead{$V$}
&\colhead{$B-V$}
&\colhead{$V-R$}
&\colhead{Vel}
&\colhead{$r$\tablenotemark{c}}
&\colhead{Category\tablenotemark{d}}
&\colhead{Class}
&\colhead{Ref.\tablenotemark{e}}
&\colhead{Comments} \\
& & & & & & & & & & 
\colhead{(km s$^{-1}$)}
& & & &
}
\startdata
J04363664-6902399 & 04 36 36.65 & -69 02 40.0 &\nodata &  9.83 & 0.46 &\nodata&\nodata&\nodata&  28.2 & 102.0 & 3 & \nodata   & \nodata               & \nodata       \\
J04364564-6858373 & 04 36 45.64 & -68 58 37.3 &\nodata & 10.93 & 0.50 &\nodata&\nodata&\nodata&  11.7 &  80.8 & 3 & \nodata   & \nodata               & \nodata       \\
J04371920-6901016 & 04 37 19.20 & -69 01 01.6 &\nodata &  8.84 & 0.71 &\nodata&\nodata&\nodata&  29.6 &  88.4 & 3 & M0        &  5  & HD 270584      \\
J04380215-6904276 & 04 38 02.15 & -69 04 27.6 &\nodata &  9.98 & 0.70 &\nodata&\nodata&\nodata&  25.6 & 140.5 & 3 & \nodata   & \nodata               & \nodata       \\
J04380671-6907112 & 04 38 06.71 & -69 07 11.3 &\nodata &  9.82 & 0.61 &\nodata&\nodata&\nodata&  31.4 & 109.1 & 3 & \nodata   & \nodata               & \nodata       \\
J04384481-6858586 & 04 38 44.82 & -68 58 58.7 &\nodata & 11.24 & 0.35 &\nodata&\nodata&\nodata& -10.8 &  94.5 & 3 & \nodata   & \nodata               & \nodata       \\
J04384847-6840267 & 04 38 48.48 & -68 40 26.7 &\nodata &  8.57 & 0.80 &\nodata&\nodata&\nodata&  75.5 &  74.7 & 3 & K7        &  5  & HD 270599      \\
J04394544-6852388 & 04 39 45.45 & -68 52 38.9 &\nodata & 10.56 & 0.55 &\nodata&\nodata&\nodata&  17.3 &  73.0 & 3 & \nodata   & \nodata               & \nodata       \\
\enddata
\tablenotetext{*}{If there is a M2002 designation, the photometry is from Massey 2002. If there is not, the photometry is from the same source of the spectral type unless otherwise noted.}
\tablenotetext{a}{These stars were imaged twice and their results were averaged.}
\tablenotetext{b}{Reference for $B-V$ colors and $V$ magnitudes = Oestreicher et al.\ 1997}
\tablenotetext{c}{Tondry \& Davis 1979 $r$ parameter.}
\tablenotetext{d}{Category: 1 = LMC supergiant; 2 = possible LMC supergiant; 3 = foreground dwarf}
\tablenotetext{e}{References for spectral classifications. 1 = Ardeberg et al.\ 1972; 2 = Bohannan \& Walborn 1989; 3 = Bohm-Vitense et al.\ 1985; 4 = Brunet et al.\ 1973; 5 = Cannon 1925; 6 = Catchpole et al.\ 1977; 7 = Cowley et al.\ 1984; 8 = Evans et al.\ 2006; 9 = Fari\~{n}a et al.\ 2009; 10 = Feast 1974; 11 = Fehrenbach et al.\ 1977; 12 = Fitzpatrick 1988; 13 = Fitzpatrick 1991; 14 = Foellmi et al.\ 2003; 15 = Gochermann et al.\ 1989; 16 = Gray \& Garrison 1989; 17 = Houk \& Cowley 1975; 18 = Humphreys 1974; 19 = Humphreys et al.\ 1991; 20 = Jaxon et al.\ 2001; 21 = Keenan \& McNeil 1989; 22 = Levesque et al.\ 2006; 23 = MacConnell \& Bidelman 1976; 24 = Massa et al.\ 2003; 25 = Massey \& Olsen 2003; 26 = Massey et al.\ 2000; 27 = Melnick 1985; 28 = Moffat \& Seggewiss 1986; 29 = Parker 1993; 30 = Rousseau et al.\ 1978; 31 = Sanduleak 1970; 32 = Sanduleak 1972; 33 = Sanduleak \& Philip 1968; 34 = Shore \& Sanduleak 1984; 35 = Stahl et al.\ 1985; 36 = Stahl et al.\ 1990; 37 = Stock et al.\ 1976; 38 = Thompson et al.\ 1982; 39 = Walborn 1977; 40 = Walborn \& Blades 1997.}
\end{deluxetable}

\begin{deluxetable}{c c c r r r r r r r r c c c l}
\tabletypesize{\tiny}
\rotate
\tablecaption{\label{tab:allobsRSG} Properties of Observed RSG Candidates\tablenotemark{*}}
\tablewidth{0pt}
\tablehead{
\colhead{2MASS}
&\colhead{$\alpha$(J2000)}
&\colhead{$\delta$(J2000)}
&\colhead{[M2002]}
&\colhead{$K_{\rm 2MASS}$}
&\colhead{$J-K_{\rm 2MASS}$}
&\colhead{$V$}
&\colhead{$B-V$}
&\colhead{$V-R$}
&\colhead{Vel}
&\colhead{$r$\tablenotemark{d}}
&\colhead{Category\tablenotemark{e}}
&\colhead{Class}
&\colhead{Ref.\tablenotemark{f}}
&\colhead{Comments} \\
& & & & & & & & & & 
\colhead{(km s$^{-1}$)}
& & & &
}
\startdata
J04434290-6746555 & 04 43 42.90 & -67 46 55.5 &\nodata &  9.86 & 1.03 &\nodata&\nodata&\nodata& 263.5 &  51.5 & 1 & \nodata   & \nodata             & \nodata       \\
J04455346-6733102 & 04 45 53.47 & -67 33 10.3 &\nodata &  9.82 & 1.11 &\nodata&\nodata&\nodata& 279.3 &  52.5 & 1 & \nodata   & \nodata             & \nodata       \\
J04464044-6711098 & 04 46 40.45 & -67 11 09.8 &\nodata &  9.62 & 1.10 &\nodata&\nodata&\nodata& 270.9 &  47.5 & 1 & B1Ia+     & 6 & HD 270754      \\
J04480007-6849112 & 04 48 00.07 & -68 49 11.3 &   3380 &  9.85 & 1.14 & 14.11 &  1.02 &  0.93 & 277.2 &  44.4 & 1 & \nodata   & \nodata             & \nodata       \\
J04480741-6922320 & 04 48 07.41 & -69 22 32.1 &   3533 &  8.21 & 1.06 & 12.30 &  1.59 &  0.92 & 122.7 &  77.0 & 3 & \nodata   & \nodata             & SP77 23-4      \\
J04481009-6945517 & 04 48 10.09 & -69 45 51.8 &   3572 &  9.92 & 0.99 & 13.63 &  1.53 &  0.81 & 285.1 &  55.5 & 1 & \nodata   & \nodata             & \nodata       \\
J04481981-6948502 & 04 48 19.81 & -69 48 50.2 &   3739 & 10.19 & 0.96 & 13.93 &  1.63 &  0.86 & 241.3 &  56.0 & 1 & \nodata   & \nodata             & \nodata       \\
J04482407-7104012 & 04 48 24.08 & -71 04 01.3 &\nodata &  9.84 & 1.01 &\nodata&\nodata&\nodata& 400.6 &  46.6 & 1 & \nodata   & \nodata             & CPD-71 285     \\
\enddata
\tablenotetext{*}{The full version of this table can be found online. If there is a M2002 designation, the photometry is from Massey 2002. If there is not, the photometry is from the same source of the spectral type unless otherwise noted.}
\tablenotetext{a}{These stars were imaged twice and their results were averaged.}
\tablenotetext{b}{These stars were previously imaged by Massey \& Olsen 2003.}
\tablenotetext{c}{Reference for $B-V$ colors and $V$ magnitudes = Oestreicher et al.\ 1997}
\tablenotetext{d}{Tondry \& Davis 1979 $r$ parameter.}
\tablenotetext{e}{Category: 1 = LMC supergiant; 2 = possible LMC supergiant; 3 = foreground dwarf} 
\tablenotetext{f}{References for spectral classifications. 1 = Ardeberg et al.\ 1972; 2 = Blanco \& McCarthy 1990; 3 = Cannon 1925;4 = Conti et al.\ 1981; 5 = Feast 1979; 6 = Fitzpatrick 1991; 7 = Henize 1956; 8 = Humphreys 1974; 9 = Hyland et al.\ 1978; 10 = Levesque et al.\ 2006; 11 = Levesque et al.\ 2007; 12 = Lundgren 1988; 13 = Massey \& Olsen 2003; 14 = Reid \& Mould 1985; 15 = Reid et al.\ 1990; 16 = Sanduleak \& Pesch 1975; 17 = Schild \& Testor 1992.}
\end{deluxetable}

\begin{deluxetable}{c c c r c c l c l l }
\tablecaption{\label{tab:knownYSG} LMC Yellow Supergiants Not Observed}
\tabletypesize{\footnotesize}
\rotate
\tablewidth{0pt}
\tablehead{
\colhead{2MASS}
&\colhead{$\alpha_{\rm 2000}$}
&\colhead{$\delta_{\rm 2000}$}
&\colhead{$K$}
&\colhead{$J-K$}
&\colhead{$V$}
&\colhead{Lit.}
&\colhead{Ref.\tablenotemark{a}}
&\colhead{Alt. Names}
&\colhead{Comment}}
\startdata
J04541425-6912366 & 04 54 14.25 & -69 12 36.7 & 7.45 & 0.57 & 10.1 & G8 0 & 3 & HD 268757; RMC 59 & Unobserved candidate \\
J04550339-6813213 & 04 55 03.39 & -68 13 21.3 & 12.21 & 0.20 & 13.2 & A9I & 5 & SOI 310; [M2002] 23123 & Unobserved candidate\\
J04560089-6626165 & 04 56 00.90 & -66 26 16.5 & 13.49 & 0.04 & 12.0 & G2Ib & 2 & RMC 63; HV 6597 & $K$ magnitude too faint\\
J04565470-6705096 & 04 56 54.70 & -67 05 09.7 & 11.37 & 0.35 & 12.7 & F0:I: & 5 & SOI 113; [M2002] 31334 & Unobserved candidate\\
J05004088-6628305 & 05 00 40.88 & -66 28 30.5 & 9.40 & 0.31 & 10.7 & F6Ia & 1 & HD 268822; Sk-66 48 & Unobserved candidate\\
J05111654-6823335 & 05 11 16.55 & -68 23 33.5 & 11.87 & 0.12 & 12.7 & F0I & 5 & SOI 368 & Unobserved candidate\\
J05124426-6715577 & 05 12 44.26 & -67 15 57.7 & 9.19 & 0.37 & 10.5 & F6Ia & 1 & HD 269154; Sk-67 60 & Unobserved candidate\\
J05210170-6548026 & 05 21 01.71 & -65 48 02.7 & 8.22 & 0.26 & 9.7 & F8 0 & 3 & HD 271182; RMC 92 & Unobserved candidate\\
\enddata
\tablenotetext{a}{References for spectral classifications and $V$ magnitudes. 1 = Ardeberg et al. 1972; 2 = Feast et al.\ 1960; 3 = Keenan \& McNeil 1989; 4 = Melnick 1985; 5 = Rousseau et al. 1978.}
\end{deluxetable}

\begin{deluxetable}{l c c c c}
\tablecaption{\label{tab:derived} Derived Properties of LMC Yellow and Red Supergiants}
\tablewidth{0pt}
\tablehead{
\colhead{2MASS}
&\colhead{Category}
&\colhead{$T_{\rm eff}$}
&\colhead{$\log{(L/L_{\sun})}$}
&\colhead{Color} \\
}
\startdata
J05055551-6753108 & 1 & 3.854 & 5.716 & Y \\
J05322496-6741536 & 1 & 3.702 & 5.600 & Y \\
J05294221-6857173 & 1 & 3.589 & 5.533 & R \\
J04551106-6710104 & 1 & 3.884 & 5.491 & Y \\
J04543683-6920221 & 1 & 3.650 & 5.478 & Y \\
J05401217-6940049 & 1 & 3.692 & 5.437 & Y \\
J05342246-6701235 & 1 & 3.891 & 5.433 & Y \\
J05051222-6644126 & 1 & 3.946 & 5.398 & Y \\
\enddata
\end{deluxetable}

\begin{deluxetable}{l r r l l r c c l l}
\tabletypesize{\footnotesize}
\rotate
\tablecaption{\label{tab:toored} Reddest Confirmed LMC Members in Our Sample}
\tablewidth{0pt}
\tablehead{
\colhead{2MASS}
&\colhead{$K$}
&\colhead{$J-K$}
&\colhead{$V$\tablenotemark{a}}
&\colhead{$V-K$}
&\colhead{[M2002]\tablenotemark{b}}
&\colhead{10$\mu$m\tablenotemark{c}}
&\colhead{Type\tablenotemark{d}}
&\colhead{Variability\tablenotemark{e}}
&\colhead{Other ID}
}
\startdata
J05364101-6914064 &   9.30 &   1.68 & 15.6\tablenotemark{f} &        6.3 & \nodata    & \nodata    & \nodata    &                    &              \\
J05365139-6913202 &  10.21 &   1.62 & 16.36      &        6.1 &  166719    & \nodata    & \nodata    &                    &              \\
J05091452-6829489 &  10.22 &   1.61 & 16.54\tablenotemark{g} &        6.3 & \nodata    & \nodata    & \nodata    &                    &              \\
J05325544-6815484 &  10.00 &   1.56 & 15.54      &        5.5 &  155787    & \nodata    & \nodata    &                    &              \\
J05253102-6616566 &   7.62 &   1.54 & 14.1\tablenotemark{f} &        6.5 & \nodata    & IMA        & \nodata    &                    &              \\
J05481398-6952301 &   9.73 &   1.51 & \nodata    & \nodata    & \nodata    & \nodata    & \nodata    &                    &              \\
J05261135-6612111 &   7.88 &   1.50 & 14.3\tablenotemark{h} &        6.4 & \nodata    & IMA        &  M3        &  slow var  (ASAS)         & WOH S 281    \\
J05381601-6910112 &   7.87 &   1.44 & 13.99      &        6.1 &  170452    &  MA        &  M4.5/5I   &  Z Dor, slow       & HTR 6        \\
\enddata
\tablenotetext{a}{$V$ magnitude is from Massey 2002 unless otherwise noted.}
\tablenotetext{b}{Catalog identification from Massey 2002.}
\tablenotetext{c}{Detection at 10$\mu$m by IRAS (``I"), MSX (``M"), and/or AKARI (``A").}
\tablenotetext{d}{Spectral type repeated from Table 1; see references there.}
\tablenotetext{e}{Variability. These are based primarily upon the OGLE data base, but the OGLE period (in days) is quoted only if the light curve appeared reasonably smooth. The CGVS and ASAS were used as indicated.}
\tablenotetext{f}{$V$ magnitude from GSC-2.3 based upon a short-$V$ plate taken with the UK Schmidt, and may be slightly brighter
than standard $V$.} 
\tablenotetext{f}{$V$ magnitude from OGLE database.}
\tablenotetext{g}{$V$ magnitude from ASAS-3.}
\tablenotetext{h}{Shorter cycles superposed.}
\tablenotetext{i}{Slow variable of 0.7~mag range.}
\tablenotetext{j}{Period confirmed in ASAS database.}
\end{deluxetable}

\begin{deluxetable}{c r r}
\tablecaption{\label{tab:lifetimes} Model Lifetimes of YSG Phase}
\tablewidth{0pt}
\tablehead{
\colhead{Mass}
&\multicolumn{2}{c}{Lifetimes ($10^3$ yr) \tablenotemark{a}} \\ \cline{2-3}
\colhead{($M_\odot$)}
&\colhead{S4}
&\colhead{S0} 
}
\startdata
12  & 147.8  & 6.3 \\
15 & 71.8  & 91.4 \\
20 & 68.4  & 50.2 \\
25 & 17.9  & 48.8 \\
32\tablenotemark{b} & 84.5 &  173.4 \\
40 & 29.9 & 21.1 \\
85 &   0    &  3.0 \\
\enddata
\tablenotetext{a}{S4 = models computed with an initial rotation of 
40\% of the critical breakup speed, S0 = models with no rotation.}
\tablenotetext{b}{The 32$M_\odot$ track represents the transition between a star ending as a RSG and blue-wards evolution. Therefore the lifetime is probably not characteristic of a models of slightly lower or higher masses.}
\end{deluxetable}

\begin{deluxetable}{c r r r r r r r r r r r r r}
\tabletypesize{\footnotesize}
\rotate
\tablecaption{\label{tab:ratios} Relative Number of YSGs Compared with Model Predictions}
\tablewidth{0pt}
\tablehead{
\colhead{}
&\multicolumn{6}{c}{S4 Models\tablenotemark{a}}
&
&\multicolumn{6}{c}{S0 Models\tablenotemark{a}} \\ \cline{2-7} \cline{9-14}
\colhead{Masses} 
&\multicolumn{2}{c}{Number}
&
&\multicolumn{3}{c}{Ratios}
&
&\multicolumn{2}{c}{Number}
&
&\multicolumn{3}{c}{Ratios} \\ \cline{2-3} \cline{5-7} \cline{9-10} \cline{12-14} 
\colhead{($M_\odot$)}
&\colhead{Class 1+2}
&\colhead{Class 1}
&
&\colhead{Class 1+2}
&\colhead{Class 1}
&\colhead{Model}
&
&\colhead{Class 1+2}
&\colhead{Class 1}
&
&\colhead{Class 1+2}
&\colhead{Class 1}
&\colhead{Model}
}
\startdata
12-15 & 25 & 24 && 1.00 & 1.00 &  1.00 && 77 & 74 && 1.00 & 1.00 & 1.00 \\
15-20 & 13 & 13 && 0.52 & 0.54 &  0.58 && 19 & 19 && 0.25 & 0.26 & 1.32 \\
20-25 &   8 &   7 &&  0.32 & 0.29 & 0.20 && 10  &  9 && 0.13 & 0.12 & 0.51 \\
(25-32)\tablenotemark{b} &  2 &   2 &&  0.08 & 0.08 & (0.19) &&   3  &  3  && 0.04 & 0.04 &(0.92)   \\
(32-40)\tablenotemark{b} &   1 &   1  &&  0.04 & 0.03 & (0.14) &&  2  &  2  &&  0.03 & 0.03 &(0.53)  \\
25-40 &   3 &   3 &&  0.12  & 0.12 &0.15 && 5   &  5 && 0.06 & 0.07 & 0.48 \\
$>$40 &  0  &  0 &&   0.00  & 0.00 &0.06 && 0   & 0 && 0.00 & 0.00 & 0.12 \\
\enddata
\tablenotetext{a}{S4 = models computed with an initial rotation of 
40\% of the critical breakup speed, S0 = models with no rotation.} 
\tablenotetext{b}{As noted earlier, the 32$M_\odot$ track represents the  transition between a star ending as a RSG and blue-wards evolution.  Therefore the lifetime is probably not characteristic of a models of slightly lower or higher masses.  We therefore suspect these bins should be ignored in evaluating the success of the models.}
\end{deluxetable}

\end{document}